# QCD EXPLORER PROPOSAL: E-LINAC VERSUS E-RING

H. Karadeniz, Turkish Atomic Energy Authority, Ankara, Turkey
S. Sultansoy, Gazi University, Ankara, Turkey

*Abstract*

TeV center of mass energy lepton-hadron collider is necessary both to clarify fundamental aspects of strong interactions and for adequate interpretation of the LHC data. Recently proposed QCD Explorer utilizes the energy advantage of the LHC proton and ion beams, which allows the usage of relatively low energy electron beam. Two options for the LHC based ep collider are posibble: construction of a new electron ring in the LHC tunnel or construction of an e-linac tangentially to the LHC. In the latter case, which seems more acceptable for a number of reasons, two options are under consideration for electron linac: the CLIC technology allows shorter linac length, whereas TESLA technology gives higher luminosity.

## INTRODUCTION

It is known that lepton-hadron collisions have been playing a crucial role in exploration of deep inside of matter. For example, the quark-parton model was originated from investigation of electron-nucleon scattering. The HERA with $\sqrt{s_{ep}} \approx 0.3$ TeV has opened a new era in this field extending the kinematics region by two orders both in high $Q^2$ and small x with respect to fixed target experiments. However, the region of sufficiently small x ($\leq 10^{-5}$) and simultaneously high $Q^2$ ($\geq 10$ GeV$^2$), where saturation of parton densities should manifest itself, is not currently achievable. The investigation of physics phenomena at extreme small x but sufficiently high $Q^2$ is very important for understanding the nature of strong interactions at all levels from nucleus to partons. At the same time, the results from lepton-hadron colliders are necessary for adequate interpretation of physics at future hadron colliders.

Today, linac-ring type machines seem to be the main way to TeV scale in lepton-hadron collisions at the constituent level (see [1] and references therein). Construction of future linear collider or a special e-linac tangentially to existing (HERA, TEVATRON, SPS, RHIC) or planned (LHC, VLHC) hadron rings will provide a number of new powerful tools in addition to ep and eA options:

- TeV scale γp [2] and γA [3] colliders. In this case high energy electron beam will be converted into photon beam using Compton back scattering of laser photons on ultra-relativistic electrons (see [4] and references therein). It should be noted that photon-hadron options can not be realized on the base of standard (ring-ring) type electron-hadron colliders. (see arguments given in [2])

- FEL-Nucleus colliders [5]. In this case (a part of) e-linac will be used for production of keV energy laser beam. Let us mentioned that FEL-Nucleus colliders satisfy all requirements on ideal photon source for nuclear resonance fluorescence experiments [6].

On the other hand, there are several standard (ring-ring type) ep collider proposals with $\sqrt{s_{ep}} > 1$ TeV. The first one is an ep option for LHC [7]. This proposal, which assumes a construction of 67.3 GeV electron ring in the LHC tunnel, is considered as a part of the LHC programme in [8]. Concerning the VLHC based ep collider, a construction of 180 GeV e-ring in the VLHC tunnel is proposed in [9]. Linac-ring type counterparts of these proposals were considered in [10] and [11], respectively.

The 2005 status was discussed in the review [12], where different TeV scale lepton-hadron and photon-hadron collider proposals (such as THERA [13], "LEP"-LHC [8, 9], QCD Explorer [1, 14] etc) are considered.

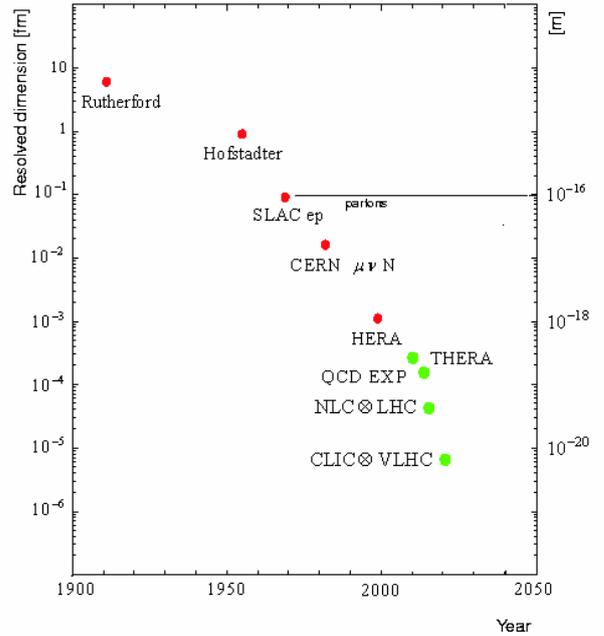

Figure 1: The development of the resolution power of the experiments exploring the inner structure of matter over time from Rutherford experiment to CLIC⊗VLHC [12].

Recently, Large Hadron Electron Collider (LHeC) is proposed, in which a 70 GeV electron (positron) beam in the LHC tunnel is in collision with one of the LHC hadron beams [15].

In this paper we compare two options for the LHC based ep collider: construction of a new electron ring in the LHC tunnel (LHeC) and construction of an e-linac tangentially to the LHC (QCD Explorer).

## QCD EXPLORER: EP OPTION

As mentioned above, reconsidered standard (ring-ring type) ep option for LHC (LHeC) assumes a construction of 70 GeV e-ring in the LHC tunnel. Main parameters of the LHeC lepton and proton beams are presented in Table 1. Center of mass energy and expected luminosity are $\sqrt{s_{ep}}$ = 1.4 TeV and $L_{ep} = 10^{33}$ cm$^{-2}$s$^{-1}$, respectively. However, construction of an additional e-ring in the LHC tunnel might cause a lot of technical problems: an example is inevitable removing of the LEP from the tunnel in order to assemble the LHC. In any case, LHC could not operate during the installation of e-ring. For these reasons, alternative linac-ring type ep option for the LHC should be considered seriously.

Table 1. Main parameters of the LHeC beams [15]

|  | Leptons | Protons |
|---|---|---|
| Beam energies, GeV | 70 | 7000 |
| Particles per bunch, $10^{10}$ | 1.04 | 17 |
| Bunch spacing, ns | 25 | 25 |
| Horizontal emittance, nm | 25.9 | 0.5 |
| Vertical emittance, nm | 5 | 0.5 |
| Horizontal β at IP, cm | 3.77 | 180 |
| Vertical β at IP, cm | 4.44 | 50 |
| Energy loss per turn, GeV | 0.676 | $6 \cdot 10^{-6}$ |
| Radiated energy, MW | 50 | 0.003 |

Let us consider the use of e-linac instead of e-ring with ~27 km circumference. The most transparent expression (in practical units) for the luminosity of linac-ring type ep colliders is [16]:

$$L = 4.8 \cdot 10^{30} \text{cm}^{-2}\text{s}^{-1} \cdot (n_p/10^{11}) \cdot (10^{-6}\text{m}/\varepsilon_p) \cdot (\gamma_p/1066)$$
$$\cdot (10\text{cm}/\beta_p) \cdot (P_e/22.6\text{MW}) \cdot (250\text{GeV}/E_e) \quad (1)$$

where $P_e$ denotes electron beam power, which is taken equal to radiation power of corresponding e-ring. With $E_e$ = 70 GeV, $P_e$ = 50 MW and LHC proton beam parameters from the Table 1 one obtain $L_{ep} = 2.4 \cdot 10^{31}$ cm$^{-2}$s$^{-1}$ for linac-ring option. If one choose the THERA proton beam parameters [13], namely, $n_p = 10^{11}$, $\varepsilon_p^N = 10^{-6}$ m and $\beta_p^*$ = 10 cm, the luminosity for "ideal" e-linac becomes $L_{ep} = 2.6 \cdot 10^{32}$ cm$^{-2}$s$^{-1}$. An additional factor 3-4 can be provided using dynamical focusing [17]. Therefore, QCD Explorer could provide for ep option the same luminosity as LHeC, in principle.

Table 2. Nominal parameters of the TESLA and CLIC

|  | TESLA | CLIC |
|---|---|---|
| Accelerating gradient, MeV/m | 23.4 | 150 |
| Bunch spacing $\tau_e$, ns | 200 | 0.66 |
| Number of bunches per pulse $n_b$ | 5600 | 154 |
| Repetition rate $f_{rep}$, Hz | 5 | 200 |
| Electrons per bunch $n_e$, $10^{10}$ | 2 | 0.4 |

Concerning the "real" e-linac technologies we consider TESLA and CLIC proposals. Parameters of the TESLA (THERA option [13]) and CLIC [18] e-beams are given in Table 2. It is seen that in the TESLA case one can use all e-bunches, whereas only ~3% of the CLIC e-bunches will collide with the LHC proton bunches. (Let us mentioned that superbunch option for the LHC could give opportunity to utilize all CLIC bunches [14] but this opportunity requires a radical modification of whole LHC stages from injector to main ring). With nominal LHC parameters we obtain $L_{ep} = 1.9 \cdot 10^{30}$ cm$^{-2}$s$^{-1}$ for "TESLA" and $L_{ep} = 1.4 \cdot 10^{28}$ cm$^{-2}$s$^{-1}$ for "CLIC" (for details see [19]). With THERA like modification of the LHC proton beam, the luminosity values become $L_{ep} = 3.3 \cdot 10^{31}$ cm$^{-2}$s$^{-1}$ and $L_{ep} = 2.3 \cdot 10^{29}$ cm$^{-2}$s$^{-1}$ respectively (see Table 3). It is seen that a factor of ~3.5 for TESLA technology and a factor of ~500 for CLIC technology are needed in order to achieve a luminosity $L_{ep} = 10^{32}$ cm$^{-2}$s$^{-1}$.

Table 3. Main parameters of "TESLA"-LHC and "CLIC"-LHC with THERA like upgrade of the LHC

|  | "TESLA"-LHC | "CLIC"-LHC |
|---|---|---|
| $n_p$ | $10^{11}$ ($5 \cdot 10^{11}$) | $10^{11}$ ($5 \cdot 10^{11}$) |
| $\beta_p$ at IP, cm | 10 | 10 |
| $\varepsilon_p^N$, μm | 1 | 1 |
| $\Delta Q_p$ | 0.0024 | 0.0005 |
| Disruption $D_e$ | 12 (60) | 12 (60) |
| $L_{ep}$, cm$^{-2}$s$^{-1}$ | $3.3 \cdot 10^{31}$ ($1.6 \cdot 10^{32}$) | $2.3 \cdot 10^{29}$ ($1.2 \cdot 10^{30}$) |

Because of 7 times higher proton beam energy comparing to the HERA the number of protons in LHC bunches can be essentially enlarged. For example, the LHC beam lifetime is ~5 h for $N_p = 5 \cdot 10^{11}$ and $\varepsilon_p^N = 1$ μm. Therefore, luminosity $L_{ep} = 10^{32}$ cm$^{-2}$s$^{-1}$ can be achieved with TESLA technology.

Radical modification of electron beam is necessary in the case of CLIC technology. For example, $N_e$ can be enlarged by the factor of 2.5 [20] (the beam-beam tune shift, $\Delta Q_p$, permits the factor ~6). In addition, the effective collision frequency can be enlarged by factor 10 due to corresponding increase of the number of bunch trains per RF pulse a la CLICHÉ [21]. Remaining factor 4 may be provided by "dynamic focusing" [17].

To summarize, using TESLA and CLIC like electron linacs with active lengths ~2.9 km and ~0.45 km, respectively, one can obtain the same center of mass energy as in the case of ~27 km electron ring. Concerning the luminosity, "moderate" upgrade of TESLA and LHC beams could give opportunity to achieve $L_{ep} = 10^{32}$ cm$^{-2}$s$^{-1}$, whereas "radical" upgrades of e-beam is needed for CLIC. Obviously, design of a "dedicated" linac will essentially improve QCD-E parameters.

## ADDITIONAL eA, γp, γA OPTIONS

Both QCD-E and LHeC can operate as eA collider with $L_{eA} \cdot A \approx 0.1 \cdot L_{ep}$, where A is the atomic number of corresponding nucleus (Pb, Cr, O, D), whereas γp and γA options are unique for QCD-E. Parameters of the QCD-E based γp collider are considered in [22] (for energy frontiers see [23]). Luminosity of γp (γA) collider depends on the distance z between conversion region and collision point and varies from ≈ 1 (at z = 0) to ≈ 0.1 (at z = 10 m) of the luminosity of basic ep (eA) collider.

## CONCLUSION

Lepton-hadron collider with $\sqrt{s_{ep}} > 1$ TeV is necessary both to clarify fundamental aspects of the QCD part of the Standard Model and for adequate interpretation of experimental data from the LHC. Today, there are two realistic proposals, namely, QCD Explorer and LHeC. Both QCD-E and LHeC will give opportunity to achieve sufficiently high luminosity to explore crucial aspects of the strong interactions. Whereas LHeC is based on the more familiar approach (we have nice experience from the HERA), QCD-E has a number of advantages:

- additional γp, γA and FEL γA options
- electron beam energy can be expanded by increasing linac length, whereas synchrotron radiation blocks this road for LHeC
- minimal influence on the LHC tunnel.

The main goal of both QCD-E and LHeC proposals is to clarify fundamental aspects of strong interactions. Their potential for the BSM physics search is restricted by center of mass energy. Therefore, very high luminosity is not so important. In our opinion γA option of the QCD-E will provide crucial information on QCD dynamics at small $x_g$ in nuclear medium

This work is partially supported by the Turkish State Planning Organization under the grant No DPT2002K-120250 and Turkish Atomic Energy Authority.